# To focus-match or not to focus-match inverse spatially offset Raman spectroscopy: a question of light penetration

GEORGINA E. SHILLITO,[1] LEWIS MCMILLAN,[1] GRAHAM D. BRUCE[1*] AND KISHAN DHOLAKIA[1,2,3]

[1]*SUPA School of Physics and Astronomy, University of St Andrews, North Haugh, St Andrews KY16 9SS, United Kingdom*
[2]*Department of Physics, College of Science, Yonsei University, Seoul 03722, Republic of Korea*
[3]*School of Biological Sciences, The University of Adelaide, Adelaide, South Australia, Australia*
*\*gdb2@st-andrews.ac.uk*

**Abstract:** The ability to identify the contents of a sealed container, without the need to extract a sample, is desirable in applications ranging from forensics to product quality control. One technique suited to this is inverse spatially offset Raman spectroscopy (ISORS) which illuminates a sample of interest with an annular beam of light and collects Raman scattering from the centre of the ring, thereby retrieving the chemical signature of the contents while suppressing signal from the container. Here we explore in detail the relative benefits of a recently developed variant of ISORS, called focus-matched ISORS. In this variant, the Fourier relationship between the annular beam and a tightly focused Bessel beam is exploited to focus the excitation light inside the sample and to match the focal point of excitation and collection optics to increase the signal from the contents without out compromising the suppression of the container signal. Using a flexible experimental setup which can realise both traditional and focus-matched ISORS, and Monte-Carlo simulations, we elucidate the relative advantages of the two techniques for a range of optical properties of sample and container.

## 1. Introduction

The use of Raman spectroscopy to perform in-situ measurements without removing a sample from its original container allows for precise chemical analysis without sample contamination or loss of value. It has therefore seen increasing application in quality assessment, product discrimination, and anti-counterfeiting in food and drink [1-3], as well as forensic [4, 5] and security applications [6, 7]. Raman spectroscopy is often conventionally performed in a back-scattering, 180º geometry, where the incident and scattered radiation are delivered and collected along the same optical path. For a sample which is contained behind an external layer, this method typically results in a bias towards signal generated from the surface. Therefore, a thick and/or highly Raman active or fluorescent barrier material can interfere with or prevent the collection of a Raman signature from the contents.

Raman spectra of the contents can be better isolated using a technique known as spatially offset Raman spectroscopy (SORS) [8, 9]. Unlike conventional Raman spectroscopy, SORS involves a lateral, physical separation of the illumination and collection sources, Δs, such that the scattered photons are collected at a distance away from the incident beam. Increasing the value of Δs allows spectra of the sample to be obtained at increasing depths. Pure sample spectra can be obtained through scaled subtraction of the spectrum acquired where the collection and excitation points overlap, i.e. Δs = 0, but in general, the signal intensity from the contents is much lower for SORS than for conventional Raman spectroscopy. Suppression of the container signal can be further improved by shifting the collection point inside the sample [10, 11]. SORS has been utilised to measure liquid and solid samples through transparent, as well as opaque containers [2, 7, 8, 12-16].

A similar technique, known as inverse-SORS (ISORS) utilises an annular beam to illuminate the sample and the scattered light is measured from the centre of the ring, such that the offset distance corresponds to the ring radius [17]. Typically, a diverging annular beam is formed using an axicon and the ring radius is varied by changing the axicon position. ISORS can give greater signal intensities compared to SORS [17, 18], but still with significantly lower signal intensity than is achievable in conventional Raman spectroscopy.

Recently, we introduced an axicon-based, back-scattering geometry to obtain Raman spectra of whisky samples, through their original glass bottles [19]. This geometry takes advantage of the Fourier relationship between the annular beam formed by an axicon and the Bessel beam generated when this is focussed by a lens. Similarly to conventional Raman spectroscopy, the Raman scattered light is collected from the excitation focal point (which is formed *inside* the sample of interest) to maximise the total Raman signal collected. Meanwhile, the incident annular beam also realises the background-suppressing properties of a spatially offset Raman spectroscopy, thereby selectively suppressing auto-fluorescence from the glass. In this work, we refer to this geometry as *focus-matched ISORS*.

What remains unclear is how the relative benefits of focus-matched and traditional ISORS are influenced by the optical properties of the sample under investigation. Here, we answer this question using a combination of a reconfigurable experimental set-up and Monte Carlo Radiation Transfer (MCRT) simulations of the light propagation [20-22], applied to a range of containers and contents.

## 2. Materials and Methods

The experiments were conducted with the arrangements shown in Figure 1. A tunable Ti:sapphire laser (Spectra-Physics 3900s) was used to excite the sample with 785 nm light, with a power of approximately 100 mW at the sample surface. The laser is sent through a single mode optical fibre (SF), collimated with a plano-convex lens (C), passed through a line filter (LF) into an axicon lens (Ax, Thorlabs Ax255-B, $\alpha = 10°$) forming a Bessel beam. The remaining lenses ($L_{1a}$, $L_{2a}$, $L_{3a}$, $L_{1b}$, $L_{2b}$ and $L_{3b}$) were achromatic doublets.

In the ISORS configuration (Figure 1a), the annular beam passes through another lens ($L_{1a}$, $f = 60$ mm) which can be moved in conjunction with the axicon lens, (Ax-$L_{1a}$, at a fixed separation distance of 133 mm) to change the point at which the Bessel beam forms, indicated by the arrows in Figure 1a. The annular beam is reflected by a dichroic mirror (DM) and onto the sample. The Raman scattering is collected in a backscattering configuration, with collection optics ($L_{2a}$, $f = 80$ mm and $L_{3a}$, $f = 80$ mm) forming a standard 4-$f$ system, passed through a notch filter (NF) to exclude Rayleigh scattering and focused into a multi-mode collection fibre (MF, 200 μm core diameter, 0.22 NA). This is passed through a spectrometer (Shamrock SR-303i, Andor Technology) onto a thermoelectically cooled CCD camera (Newton, Andor Technology).

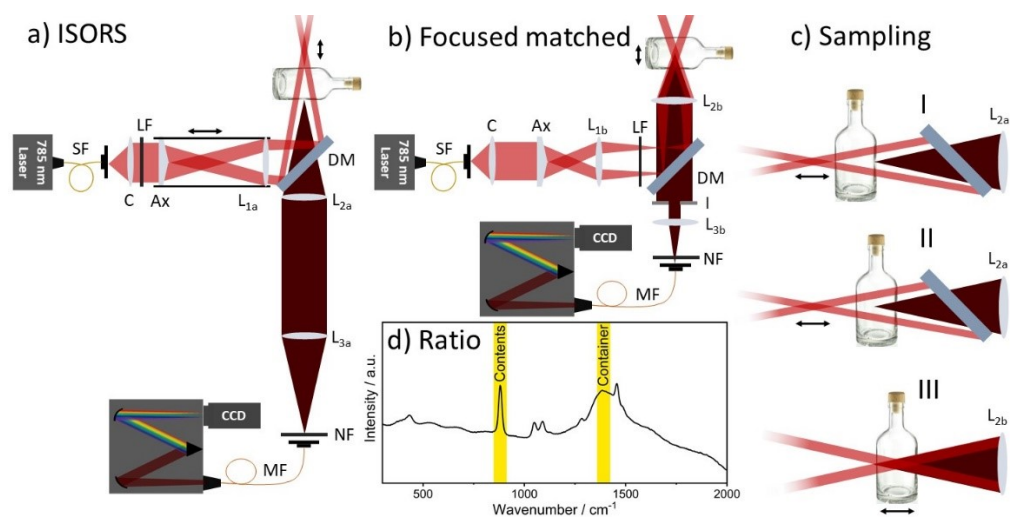

Fig. 1. Experimental geometries. a) Flexible ISORS configuration, where the axicon (Ax) and first lens ($L_{1a}$, $f = 60$ mm) can be moved as a unit along the optical cage, varying the position of the second Bessel beam and hence the ring radius on the sample surface. b) The focus-matched geometry, where the excitation and collection pathways overlap through $L_{2b}$ ($f = 40$ mm), and the ring size can be varied through moving the position of the sample. c) The sampling configurations: (I) ISORS configuration, with collection from the surface; (II) ISORS configuration, with collection inside the sample; (III) overlapping collection and excitation for the focused-matched configuration.

In the focus-matched configuration (Figure 1b), the axicon and $L_{1b}$ ($f = 100$ mm) are in a fixed position 113 mm apart. The excitation and collection geometries overlap through a lens ($L_{2b}$, $f = 40$ mm) placed after the dichroic mirror (DM), which forms part of the 4-$f$ collection system ($L_{3b}$, $f = 50$ mm). An iris (I) with a diameter of ~10 mm, is placed between $L_{2b}$ and $L_{3b}$ which functions to help exclude Raman signal from the sample surface, such that the majority of photons are collected from inside the ring. The use of different lens combinations in the focus-matched system was explored, with further details found in the Supporting Information (Figure S1).

In the ISORS configuration, Figure 1a, movement of the Ax-$L_{1a}$ combination varies the position of the Bessel beam, and therefore the ring radius on the sample surface. As shown in Figure 1c, the collection position was either directly on the surface of the sample (I) or at a fixed distance inside the sample, beyond the container surface (II). The collection point was altered by adjusting the position of the sample. The flexibility of movement of the ISORS set-up also allows the realisation of both a conventional Raman geometry (where the collection and excitation occur from a spot, $\Delta s = 0$ mm, on the sample surface) and the focus-matched configuration (where the collection and excitation points overlap inside the sample). In the separate, focus-matched configuration, Figure 1b, the excitation and collection always overlap, so the ring radius and hence the focus and collection point is varied through movement of the sample (III).

We investigated the benefits of the focus-matched approach in a range of samples. Samples measured experimentally range from almost-transparent to highly opaque. These include whiskies in their original clear glass bottles with a diameter of 70 or 35 mm as well as a white, opaque plastic bottle made of high density polyethylene (HDPE) filled with paracetamol tablets available over-the-counter. In addition, a white Lego block, made from 5 mm thick acrylonitrile butadiene styrene (ABS), concealing a crushed paracetamol tablets was also measured. Spectra were obtained using a 2 s exposure time and 30 accumulations. The container-to-contents ratio, with intensities measured from the baseline was calculated.

Here a *decrease* in value corresponds to an *increase* in relative contribution of the contents signal. In certain situations (*vide infra*) the contents signal is excluded entirely, leading to a ratio of zero. Error bars represent the standard deviation between three repeat measurements where the sample was removed and replaced.

All measurements had a dark (CCD slit closed) background correction. For the plastic containers, a baseline subtraction was performed to remove a portion of the background to better compare relative Raman intensities of the container and contents bands. In the case of the alcohols in glass bottles no baseline subtraction was performed. The ISORS spectra are presented without any scaled subtraction of a container reference (i.e. $\Delta s = 0$) spectrum for all sample types. Signal-to-noise (SNR) calculations were performed using the method reported by McCreery [23]. The SNR was calculated by taking the mean peak height above the baseline divided by σ. The value of σ was determined by subtraction of two successive spectra and taking the standard deviation of the resulting noise spectrum in the peak region, divided by √2. Spectral processing was performed using SpectrGryph software [24].

To establish the range of samples for which the focus-matched geometry offers an advantage, and in particular to investigate the parameter range between experimentally measured samples in a controllable way, we performed numerical simulations of the light propagation inside the media of interest. As the interaction of light and turbid media is not easily elucidated, we employ simulations of light transport using the MCRT method to shine light on this problem. MCRT is a stochastic method that uses interaction probabilities and random numbers to fully simulate the propagation of light though various media. We employ it in this work to determine the evolution of the excitation intensity within the container, and thus determine in what types of media our focus-matched ISORS geometry excels.

The MCRT code we used in this work was originally developed for simulating light transport in dusty galaxies [22], and has since been adapted for use in medical [20], and biophotonics [21] applications. Here, we specifically use the newly-developed signedMCRT method, which allows the accurate modelling of arbitrary curved surfaces [25]. Light impinges on a semi-infinite slab, whose optical properties (specifically the scattering and absorption coefficients, refractive index and the anisotropic scattering parameter) can be set such that they mimic different media of interest. The medium is divided into 200 x 200 x 200 voxels giving a resolution of 0.15 mm and the propagation of $10^7$ photons is simulated, with an initial position- and direction-distribution chosen to match the experimental geometry. The number of voxels gives sufficient resolution for the problem, and the number of photons simulated gives an adequate signal to noise ratio for the mean intensity. The trajectory of each photon is traced to the desired focal plane, as is the backward propagation of photons to the plane of the collection fibre. To parameterise the effect of the medium properties, we compare the light intensity reaching the fibre plane (taking only photons within the central 200 mm and an angle of incidence within the typical numerical aperture (0.22) to mimic the fibre collection) from photons originating both within the sample of interest and from the container. Simulations were fully parallelised and run on an AMD Ryzen 9 3950X 16-Core Processor and took between 0.5 and 3.0 minutes per sample.

## 3.  Results and Discussion

### 3.1 ISORS optimization in opaque and transparent samples

As mentioned previously, typical ISORS measurements are performed by scaled subtraction of a zero-offset spectrum from that obtained at a specific radial offset. The zero-offset measurement is performed when the annular beam is focussed to form a spot on the sample's surface. When the collection point is also on the sample's surface, overlapping with the excitation, the resulting geometry is analogous to a conventional Raman configuration, such that the majority of the signal contribution is due to the container. In general, the greater the radial offset distance, the greater the relative contribution of photons from sub-surface contents to the spectrum [26]. Such a behaviour was reported by Matousek for a two-layer system of trans-stilbene powder contained behind a layer of PMMA powder, where increasing the ring radius from 0.9 to 3.8 mm reduced the surface layer signal by a factor of ~13, but was also accompanied by a decrease in overall signal strength [17]. However, as shown in Figure 2a, this typical behaviour is not observed for the ISORS measurements taken of whisky samples in clear glass bottles using the experimental setup in Figure 1a. Figure 1c (I and II) illustrates how the sample position was adjusted such that spectra could be obtained with the collection focus fixed either at the sample surface (blue lines in Figure 2) or at a set distance inside the sample (red lines in Figure 2). Movement of the Ax-$L_{1a}$ unit varies the ring radius on the sample surface as well as the position of formation of the Bessel beam, moving from the surface of the bottle through, and in some cases, beyond the back of the sample.

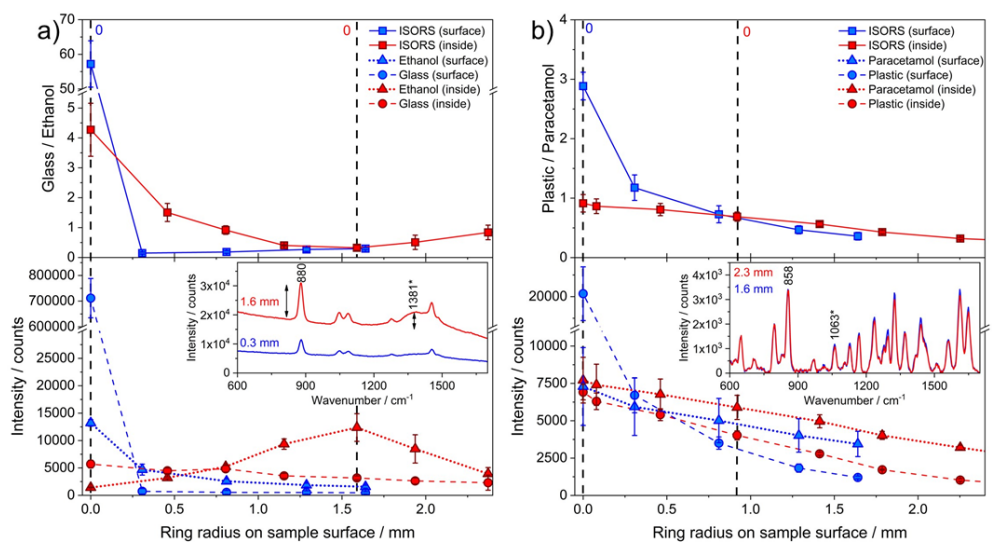

Fig. 2. Focus-matched ISORS achieves comparable glass extinction to traditional ISORS, but gives increased signal from the contents for (a) whisky in glass bottle, but offers no advantage for (b) paracetamol in plastic. Upper panels: The relationship between the ring radius on the sample surface and the container-to-contents ratio. Lower panels: the intensities of the respective container and contents bands as a function of beam ring radius. Dashed vertical lines indicate the 'zero offset' positions, where the excitation and collection paths overlap, giving an approximate conventional Raman geometry (surface) and a focus-matched geometry (inside). Lower panels show the intensities of the respective container and contents bands as a function of ring radius. The insets show spectra obtained at their lowest container-to-contents ratio, container band marked with an *. Arrows indicate method of measurement of peak heights.

As shown in the upper panel of Figure 2a, the ratio between the relative intensity of the fluorescence signal from the glass at 1381 cm$^{-1}$ and the Raman peak from ethanol at 880 cm$^{-1}$ (inset Figure 2) was plotted against the internal radius of the ring on the sample surface. As anticipated, the highest ratio (with the greatest surface contribution) of ~ 60 occurs in the conventional Raman configuration, when signal is collected from the surface of the sample and the ring has a radius of 0 mm. The signal level and the container-to-contents ratio obtained with traditional ISORS does not show a strong dependence on the ring radius, only requiring that the excitation does indeed have an intensity minimum on axis, through which the Raman signal of the contents can be collected.

When the collection point is moved to 22 mm inside the bottle, parabolic-like behaviour of the container-to-contents ratio as a function of ring radius is observed, where the best suppression of the glass, a ratio of 0.32, is obtained with a radius of 1.6 mm. At this point, the Bessel beam forms inside the sample and overlaps with the collection point and hence corresponds to the 'focus-matched' configuration. While the suppression of the glass is slightly poorer than with the surface collection, there is a considerable improvement made in the overall spectral intensity, particularly that of the ethanol signal, being approximately 3x that obtained at the best glass suppression point with surface collection (inset Figure 2a). This increase in signal strength is accompanied by a similar improvement in the SNR of the 880 cm$^{-1}$ band, from 70 in the traditional ISORS geometry to 180 in the focus-matched ISORS geometry. This increase in intensity of the ethanol signal is illustrated in the lower panel of Figure 2a, reaching its maximum at the focused-matched configuration (dashed line). This focus-matched geometry simultaneously realises the advantages of both conventional Raman spectroscopy and conventional ISORS, achieving a signal intensity from the contents which is comparable to that achieved with conventional Raman, with an ability to suppress the surface signal which is comparable to that achieved with ISORS (see also Figure S2). The potential application of ISORS in a focus-matched geometry was therefore investigated further.

The same experiment was repeated with paracetamol tablets inside an opaque white HDPE bottle. As shown in Figure 2b, for both surface and internal collection, as the ring radius increases, the plastic-to-paracetamol ratio (measured at 1063 and 858 cm$^{-1}$, respectively) decreases, corresponding to a relative increase in signal contribution from paracetamol and hence exhibits typical ISORS behaviour. Interestingly, when the excitation beam is focused on the surface, an improvement in the container-to-contents ratio without sacrifice in contents signal is obtained by applying an axial offset [10, 11], i.e. moving the collection point 15mm inside the bottle. At a radius of 0.9 mm, the conditions for a focus-matched configuration are achieved. However, unlike with the whisky samples, no notable change in the container-to-contents ratio is seen at this point. This is also the case for a Lego block concealing a layer of paracetamol powder - further details can be found in Figure S3. To better understand why the focus-matched geometry does not outperform typical ISORS behaviour in more highly scattering media, we undertook a numerical study using MCRT to simulate the propagation of light in the various scattering media.

### 3.2 The role of sample optical properties

To investigate under which conditions the use of the novel focus-matched ISORS geometry is advantageous, we numerically modelled the propagation of the beam in a semi-infinite slab with variable scattering and absorption properties, a fixed refractive index (1.3), and anisotropy g value (0.7) [27] and propagated an annular beam into a focus at 1.55 cm depth. Figure 3 shows the results of these MCRT simulations.

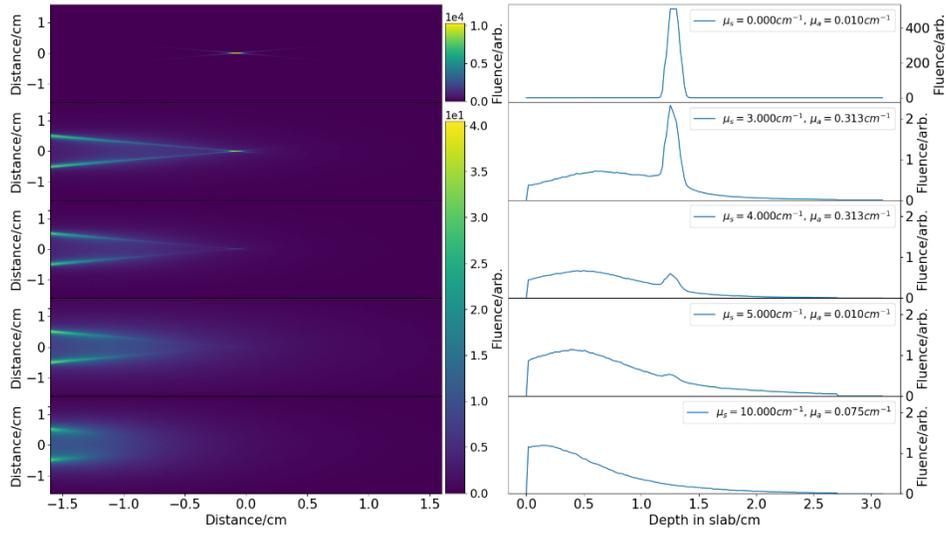

Fig. 3. Penetration of excitation light into samples with differing optical properties. Left panels: Cross section of intensity on the y-z plane at x=0 for a range of optical properties, $\mu_s$ =[0, 10] cm$^{-1}$ $\mu_a$=[0.01,0.313] cm$^{-1}$. As the optical properties are increased, we see that the annular beam no longer focuses to a point. Right panels: Average intensity of the 9 pixels around the line x=y=0. As we increase the scattering and absorption coefficients of the slab, ballistic photons are prevented from reaching the focal distance (peaks in top 3 panels). At higher scattering coefficients, scattering dominates and produces the characteristic scattering curve near the surface of the slab.

The Raman signal from the contents will be directly proportional to the peak intensity in the medium. Figure 3 shows the peak intensity along the central axis of the beam, for varying scattering and absorption coefficients. In the high-penetrations limit (no scattering or absorption), the intensity shows a large peak at the focal location of the annular beam. The peaks at ~1.4 cm correspond to the offset focal depth due to the refractive index at the boundary and comes from ballistic photons which are unscattered by the media. As the penetration depth is reduced, the height of the peak is lowered whilst a second broader peak appears at shallower depth. This shallower peak is the peak of the scattered photons as opposed to the ballistic photons. Figure 3 shows clearly that as the penetration depth is lowered the ballistic peak becomes negligible compared to that of the scattered peak. The focus-matched ISORS geometry is critically dependent on an abundance of ballistic photons, i.e. photons that undergo little to no scattering, to produce the strong signal from the contents. Indeed, this is the key difference between ISORS and the focus-matched ISORS, in that for traditional ISORS ballistic photons do not impinge on the optical axis within the medium, as shown in Fig 1c (I). In order for focus-matched ISORS to provide an improvement over traditional ISORS, the optical penetration depth of the medium must be sufficiently high that a tightly-focussed spot can be formed near the centre of the container by ballistic photons. This is why there is no marked difference between the ISORS geometries for highly scattering media, while the focus-matched configuration provides an additional signal boost in weakly scattering media. For the highly scattering case, a slight boost in contents signal can be achieved, if the focal depth is carefully selected. As seen in Figure S4, there is an empirical relationship between the optical penetration depth and the 1/e point of the peak intensity. Setting the focal point of the focus-matched geometry no deeper than this depth will give a slight boost to the contents signal.

For illustrative clarity, we have given the above demonstration for a flat container. In many samples of interest, containers will have curved surfaces, which will create astigmatism or other aberrations in the beam. However, for typical glass bottles, this astigmatism does not prevent the focus-matched ISORS approach from improving the signal and contents-container ratio, as demonstrated in Figure S5.

*3.3 Optimising focus-matched ISORS*

The set-up in Figure 1b was used to study the same whisky samples in an optimised focus-matched ISORS configuration. In order to achieve the best surface suppression effect and signal intensity, a judicious choice of lenses can be made.

Most obvious of these, is that the lens $L_{2b}$ should be chosen with the highest possible numerical aperture (shortest focal length) to optimise the signal collection. Additionally, especially for smaller bottles, setting the collection point in the centre of the container gives optimal glass suppression (see Figure 4). This is because maximising the offset distance from any glass surface minimizes its contribution to the total signal, and moving the collection point beyond the centre of the bottle causes signal from the back surface of the container to be detected (which we verified in our MCRT model, Figure S6). Combining the optimal glass suppression distance with the requirement for highest signal collection efficiency, we see that the optimal focal length of $L_{2b}$ is approximately equal to the bottle radius.

Increasing the focal length of $L_{1b}$ increases the demagnification of the Bessel beam produced by the axicon. This minimises the size of excitation spot, producing a high intensity and thus high signal from the contents. This also increases the size of the ring entering $L_{2b}$, and thus the ring size on the bottle surface, reducing the signal from the container which is collected by the fibre. The ratio $L_{3b} / L_{2b}$ can be chosen to maximise the collection of the contents signal while simultaneously ensuring that the signal from the glass surface misses the fibre front facet, as demonstrated in Figure S7.

Figure S1 shows the suppression effect on a whisky sample for four different combinations of lenses with varying focal lengths. For 70 mm-diameter glass bottles, we find an optimal combination of focal lengths to be $L_{1b}$ = 100 mm, $L_{2b}$ = 40 mm, $L_{3b}$ = 50 mm.

In our previous work [19] it was noted that inclusion of an iris aperture functioned to further exclude fluorescence signal from the glass bottle, whilst selectively retaining signal from the alcohol contents. In the optimised focus-matched geometry, the role of the iris aperture was explored in more detail. In the case of the 70 mm diameter whisky bottle in Figure 4a, inclusion of the iris considerably improves the glass suppression effect over a broad range of ring sizes, such that a maximal suppression is achieved for a range of bottle positions, as long as the Bessel beam is formed inside the approximate middle region of the sample. In fact, as illustrated in the insets of Figure 4, inclusion of an iris can suppress the glass signal completely. The resulting spectra obtained are free of the glass signal and almost exactly match that obtained from the same whisky sample measured inside a quartz cuvette (Figure S8). The glass-to-ethanol ratio therefore reaches zero when no glass signal is observed.

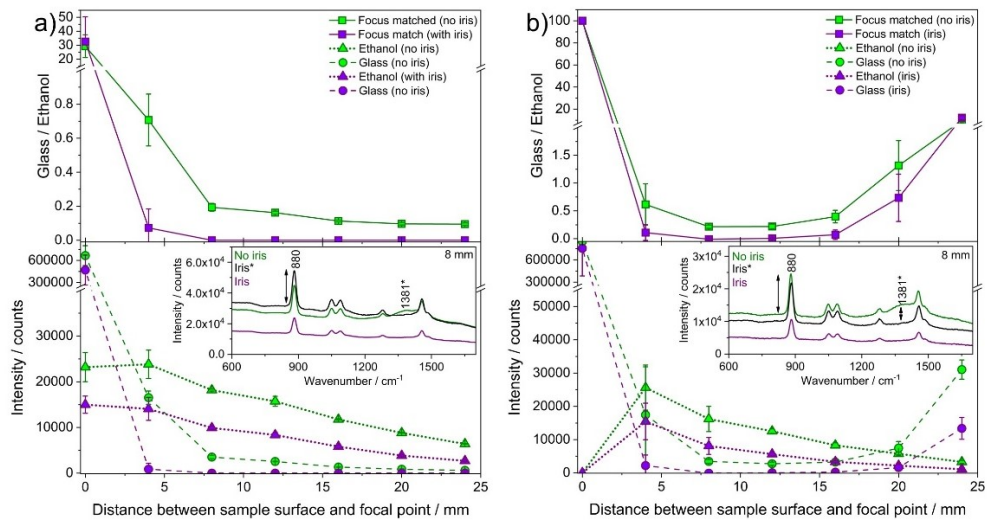

Fig. 4. The effect of inclusion of an iris in the exclusively focus matched configuration, demonstrating that the glass signal can be excluded in entirety. Upper panels: The effect of the separation distance between the focal point (collection and excitation) of $L_{2b}$ and the sample surface on the container-to-contents ratio for whisky samples in clear glass bottles of a) 70 and b) 35 mm diameter. Lower panels show the intensities of the glass and ethanol bands as a function of separation distance. The insets show spectra obtained at their lowest container-to-contents ratio with container peaks labelled with an *. The green and purple spectra correspond to spectra taken under identical conditions. The black spectrum* is obtained with an iris but with an increased CCD slit width to improve signal intensity. Arrows indicate method of measurement of peak heights.

We note that the optimal position for an iris would intuitively be on an image plane of the excitation focus, realising a confocal pinhole geometry, which is well known to exclude out-of-plane light. However, given the high degree of extinction demonstrated without this positioning for the case of the whisky bottles, in the interest of simplicity we have elected not to add the extra elements required to realise this additional image plane. Inclusion of the iris does decrease the overall signal intensity, however the spectra are still of good quality and can be further improved by increasing the size of the entrance slit to the CCD (Figure 4 insets). We find that an aperture diameter of around 10 mm is a good compromise between signal level and glass suppression: smaller apertures significantly reduce the total signal without further improving the container-to-contents ratio (Figure S9). This focus-matched configuration, combined with an iris aperture, allows for the acquisition of container-free spectra of a glass bottle's contents without the need for the typical scaled spectral subtraction performed in traditional ISORS.

## 4. Conclusion

In the case of weakly scattering media, adoption of a focus-matched ISORS configuration where the excitation and collection foci overlap presents several advantages over the conventional ISORS set-up. Namely, the ability of photons to penetrate ballistically deep inside the sample, and therefore form a tight focus, leads to an increased generation of Raman scattered photons from the contents. This significantly boosts the signal of contents relative to that of the container. Furthermore, photons collected from the container can be further suppressed by inclusion of an iris aperture in the system. In the case of whisky in a clear glass bottle, we have demonstrated that the utilisation of this method allows the fluorescence signal of the glass to be bypassed entirely, facilitating acquisition of isolated Raman spectra of the contents, without the need for spectral subtraction. Larger diameter bottles allow more flexibility in the sample's positioning, due to reduced contribution from the back glass wall.

The focus-matched configuration does not hold any notable advantages for highly scattering media, such as opaque plastic and paracetamol tablets, due to the limited penetration depth. However, importantly, offsetting the collection focus beyond that of the sample's surface can help to reduce signal attributed to the container [10, 11].

In conclusion, we have shown that when designing an offset Raman spectroscopic probe for samples inside containers, one should pay careful attention to the scattering and absorption properties of the container and the medium, in order to optimally detect the contents while minimising signal from the container. When the optical penetration depth of the sample is larger than or approximately equal to the depth at which the sample is probed, the tightly-focussed beam produced in the focus-matched ISORS configuration will offer an enhancement of the signal of the contents while maintaining the container extinction properties of traditional ISORS. For example, the signal of the contents for whisky in original glass containers is enhanced by a factor of 3 in both signal strength and signal to noise ratio. The focus-matched geometry should prove to be particularly useful for anti-counterfeiting and adulterant detection of liquids, and has the potential to function as a compact, portable device.

**Funding.**

The work was supported by funding from the UK Engineering and Physical Sciences Research Council (EP/P030017/1 and EP/R004854/1) and the H2020 FETOPEN project "Dynamic" (EC-GA 863203).

**Acknowledgements.**

The authors thank Mingzhou Chen for useful discussions and Rory M. Bruce for the loan of Lego used in this study

**Disclosures.**

The authors declare that there are no conflicts of interest related to this article

**Data availability**

The data that support the findings of this study will be openly available via the University of St Andrews Open Data Repository

**Supplemental document.**

See Supplement 1 for supporting content.

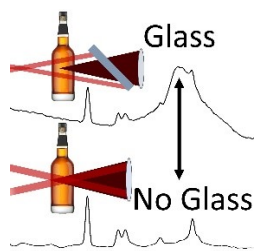

# To focus-match or not to focus-match inverse spatially offset Raman spectroscopy: a question of light penetration
# : supplemental document

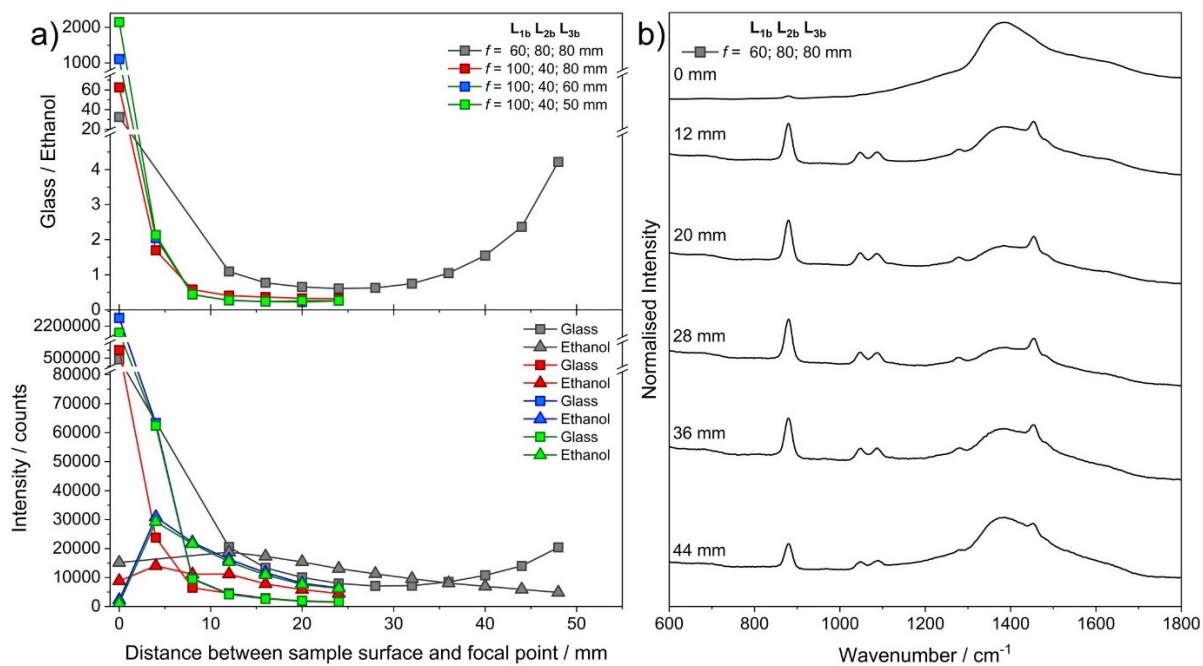

Fig. S1. The effect of different lens combinations in the focus-matched geometry for a clear glass 700 mL whisky bottle of 70 mm diameter. a) The glass-to-ethanol ratio. The position of the bottle was moved from $L_{2b}$ to just beyond the point of Bessel beam formation (where the separation between the sample surface and the focal point is 0 mm), and the corresponding separation distances measured. The change in ratio was measured by comparing the glass peaks and ethanol peaks at 880 and 1381 cm$^{-1}$. b) The resulting spectra for $L_{1b}$ = 60, $L_{2b}$ = 80, $L_{3b}$ =80 mm.

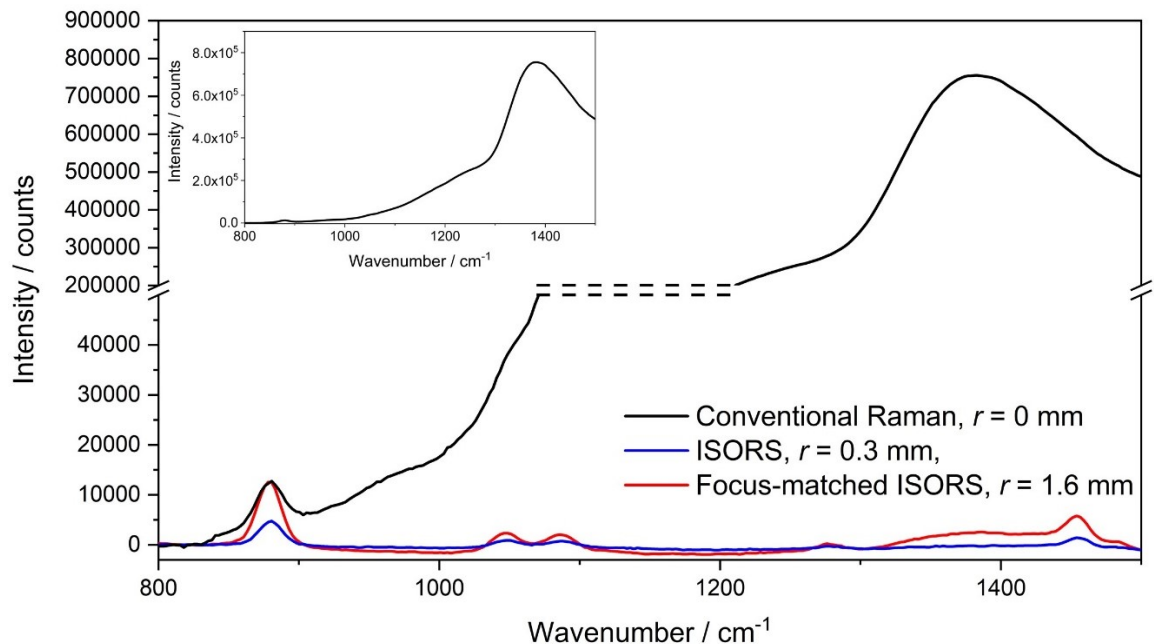

Fig. S2. Spectra of whisky in a 70 mm diameter bottle. The black spectrum is obtained with collection focus on the surface of the glass and a ring radius of 0 mm and is hence analogous to a conventional Raman spectrum, predominantly giving rise to signal from the glass bottle. The blue curve shows the spectrum obtained with collection on the surface of the bottle and a ring radius of 0.3 mm, which gives the best glass-to-ethanol ratio. The red spectrum shows the result obtained in the focus-matched configuration, where collection and excitation points overlap inside of the bottle. Spectra have had a linear baseline correction performed at 800 cm$^{-1}$ so that the height of the 880 cm$^{-1}$ relative to the fluorescence background can be more easily compared. No scaled SORS subtractions have been performed. The inset shows the whole conventional Raman spectrum, without the y-axis break.

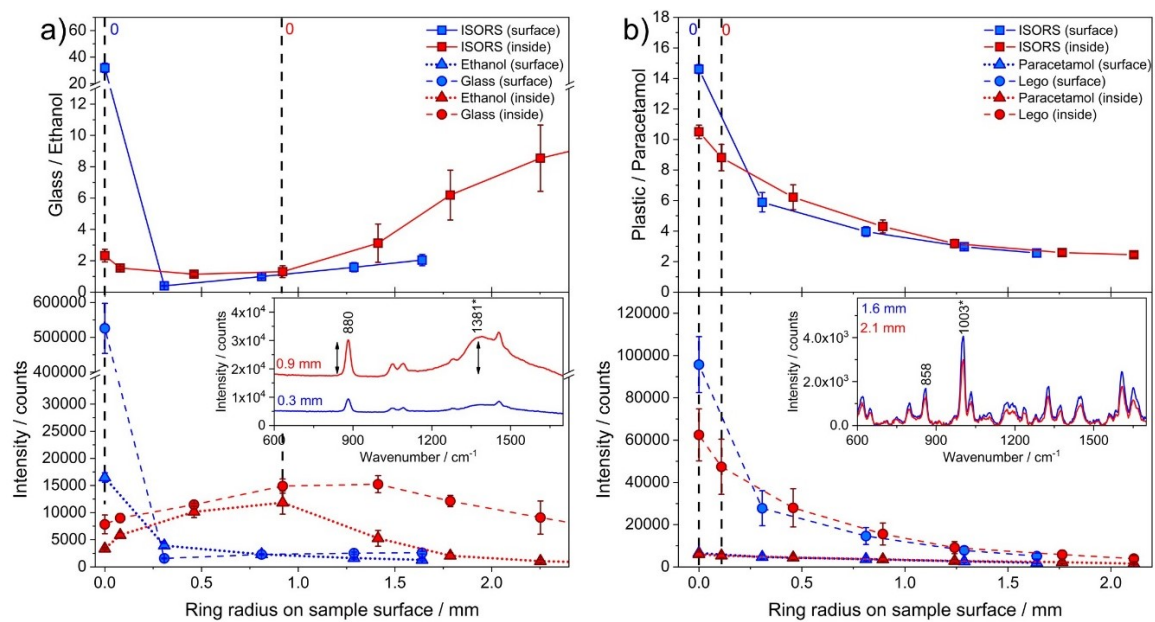

Fig. S3. Focus-matched ISORS achieves comparable glass extinction to traditional ISORS, but gives increased signal from the contents for (a) whisky in 35 mm diameter glass bottle, but offers no advantage for (b) crushed paracetamol behind a 5 mm white Lego (ABS plastic) block. Upper panels: The relationship between the ring radius on the sample surface and the container-to-contents ratio. Dashed vertical lines indicate the 'zero offset' positions, where the excitation and collection paths overlap, giving an approximate conventional Raman geometry (surface) and a focus-matched geometry (inside). Lower panels: The intensities of the respective container and contents bands as a function of ring radius. The insets show spectra obtained at their lowest container-to-contents ratio, container band is marked with an *. Arrows indicate the method of measurement of peak heights.

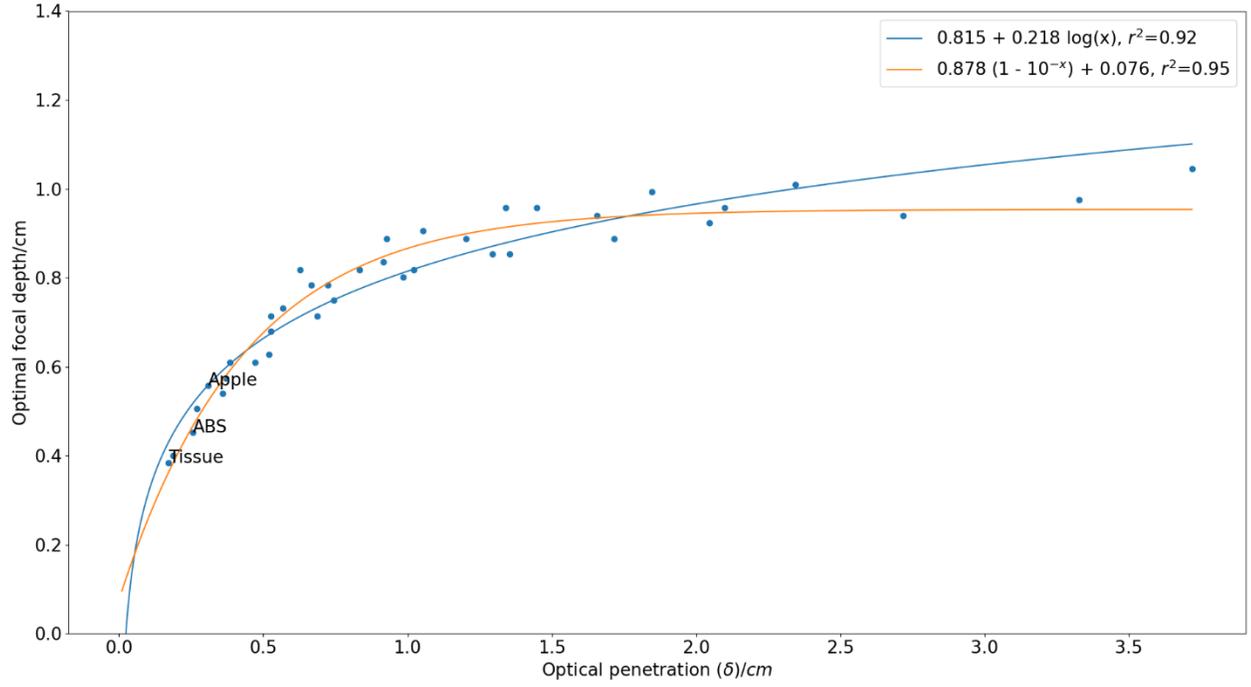

Fig. S4. Optimal focal depth for a slab with highly scattering optical properties ($\mu_s > 5$ cm$^{-1}$). Focal depth is the 1/e of the peak intensity along the x=y=0 line. Optical penetration depth defined as $\left(\sqrt{3\mu_a(\mu_a + \mu_s(1-g))}\right)^{-1}$

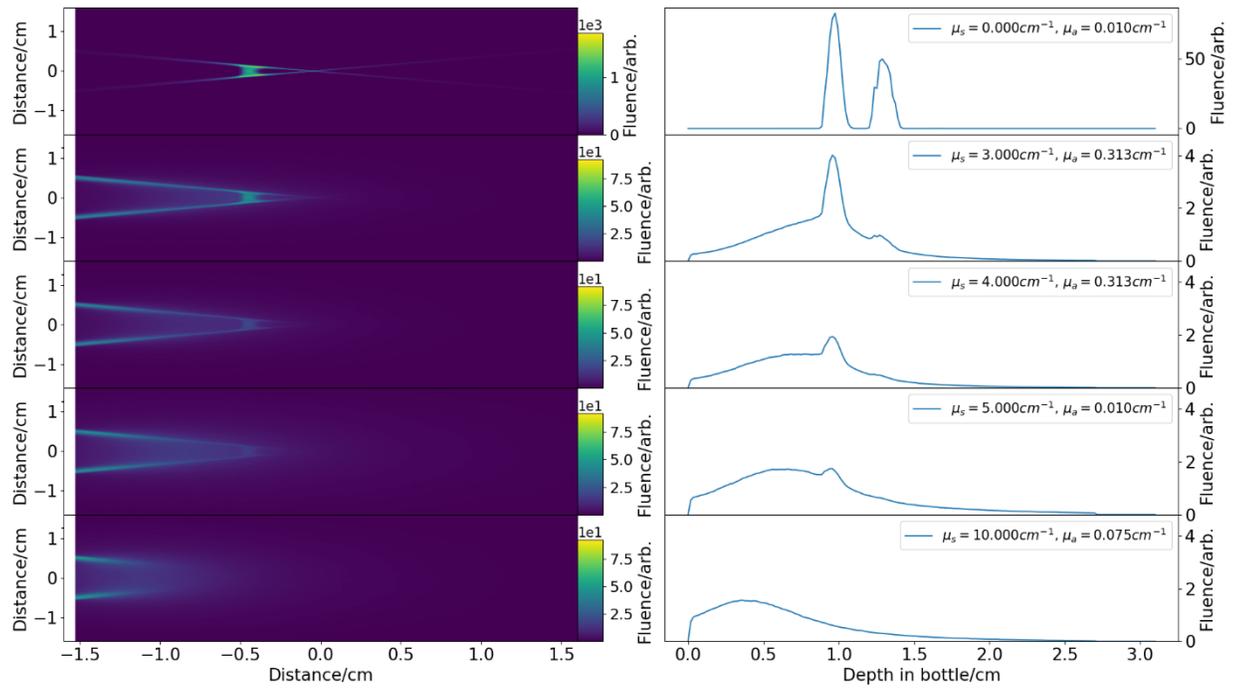

Fig. S5. Penetration of excitation light into a bottle for samples with differing optical properties. Left panels: Cross section of intensity on the y-z plane at x =0 for a range of optical properties, $\mu_s$ = [0,10]cm$^{-1}$ $\mu_a$=[0.01, 0.313]cm$^{-1}$. As the optical properties are increases, we see that the annular beam no longer focuses to a point. Right panels: Average intensity of the 9 pixels around the line x=y=0. As we increase the scattering and absorption coefficients, scattering dominates and produces the characteristic scattering curve near the surface of the bottle.

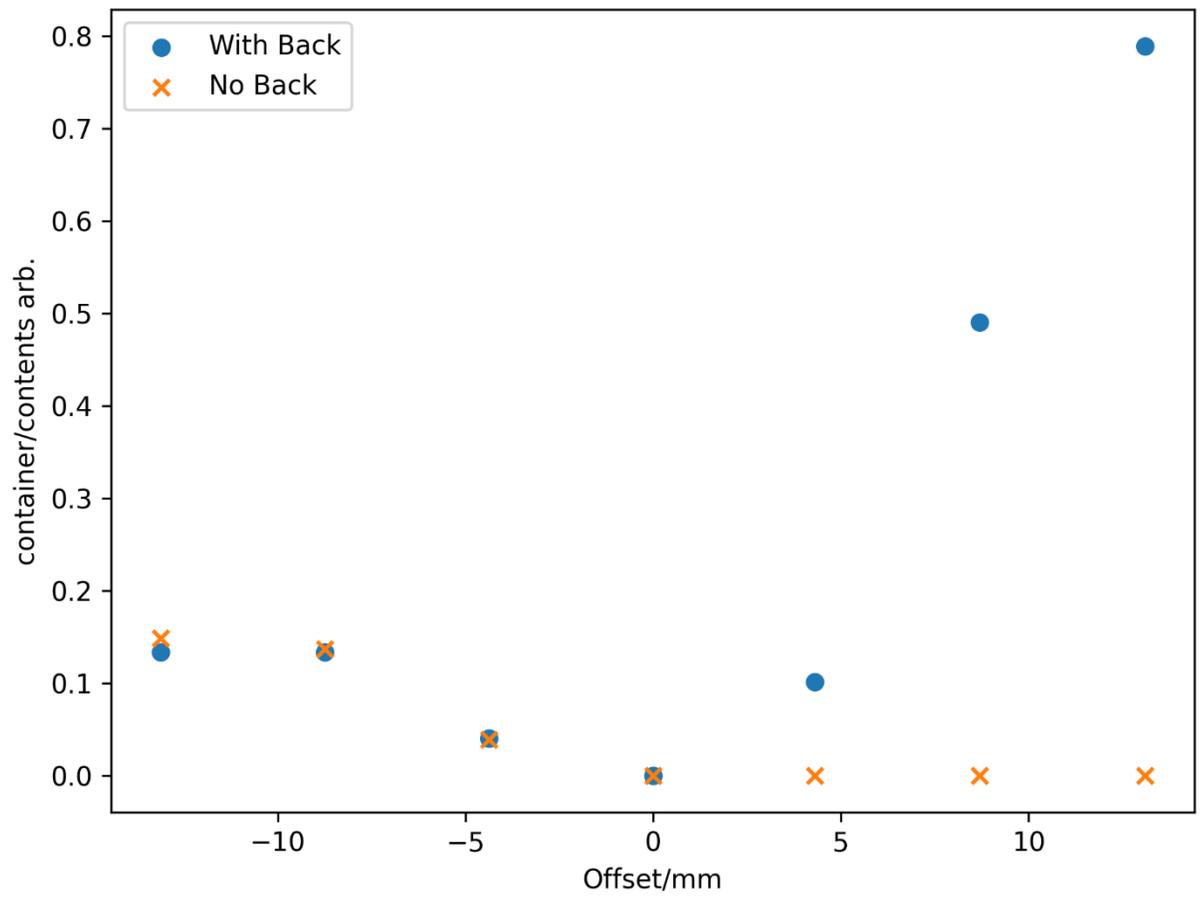

Fig. S6. Monte Carlo radiation transport simulation of light from bottle through optical setup. Comparison of bottle with no glass back and glass back.

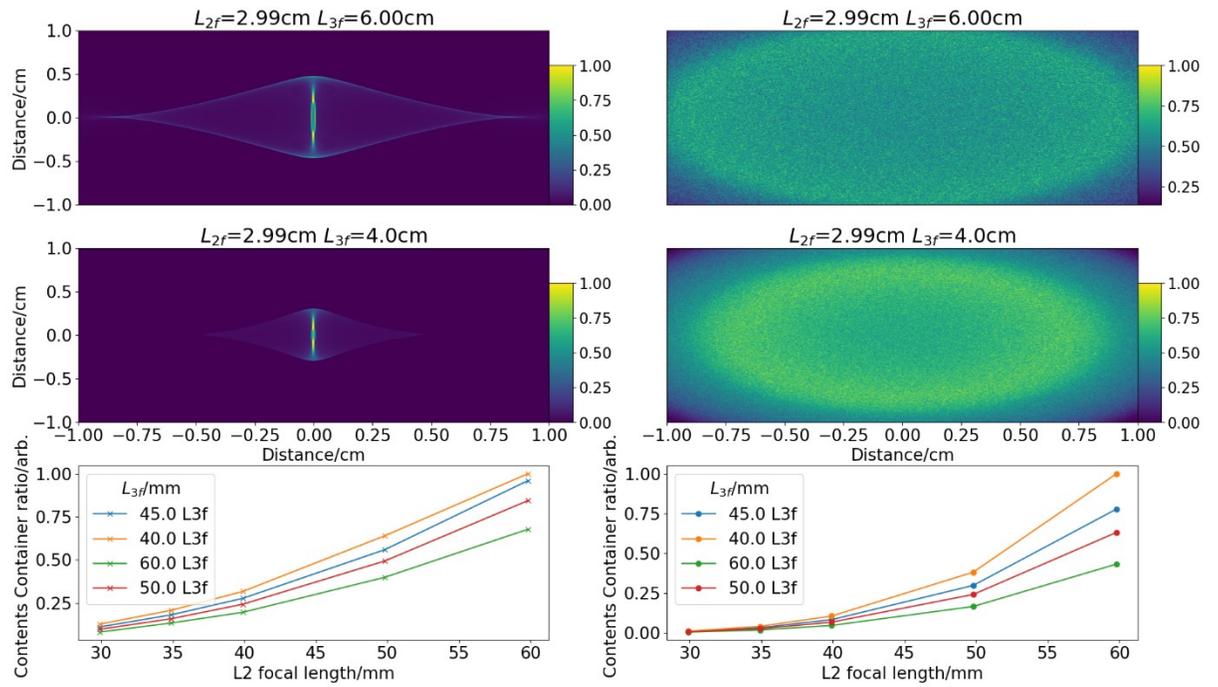

Fig. S7. Monte Carlo radiation transport simulation of optical setup for different L2 and L3 lenses. Top four panels show the image at the fibre plane from the contents (left two panels), and container (right two panels). Bottom panel shows the log ratio of contents to container intensity as measured by the simulated fibre. Equal amounts of photons were released from the contents (modelled as a point source) and container (modelled as a ring of point sources) sources. From the bottom panel it can clearly be see that the choice of L2 has more of an effect than the choice of L3.

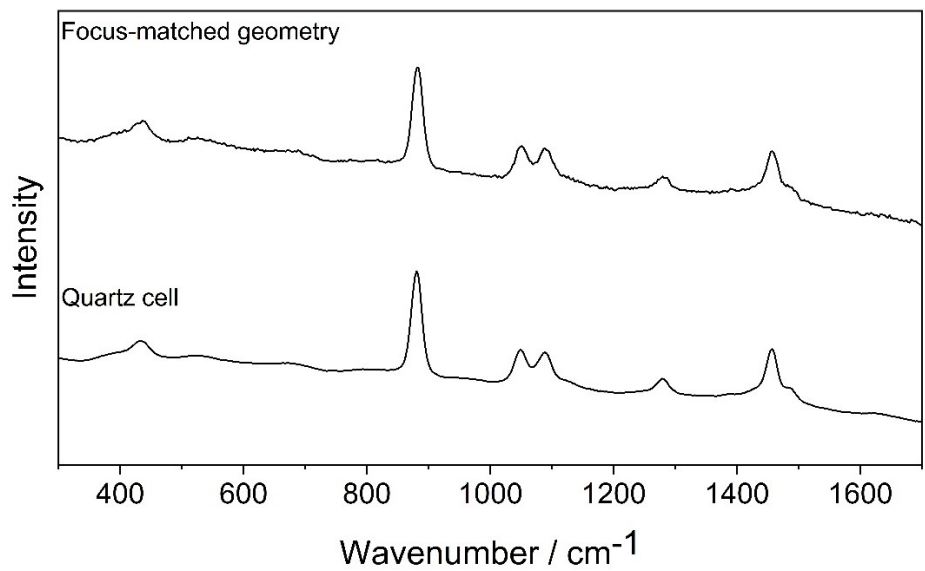

Fig. S8. Focus-matched ISORS spectra of a whisky sample in its original clear glass bottle, compared to that of the same whisky which was extracted and measured in a quartz cuvette. Spectra are offset for clarity. It can be clearly seen that the glass signal is excluded without the need to perform a scaled subtraction.

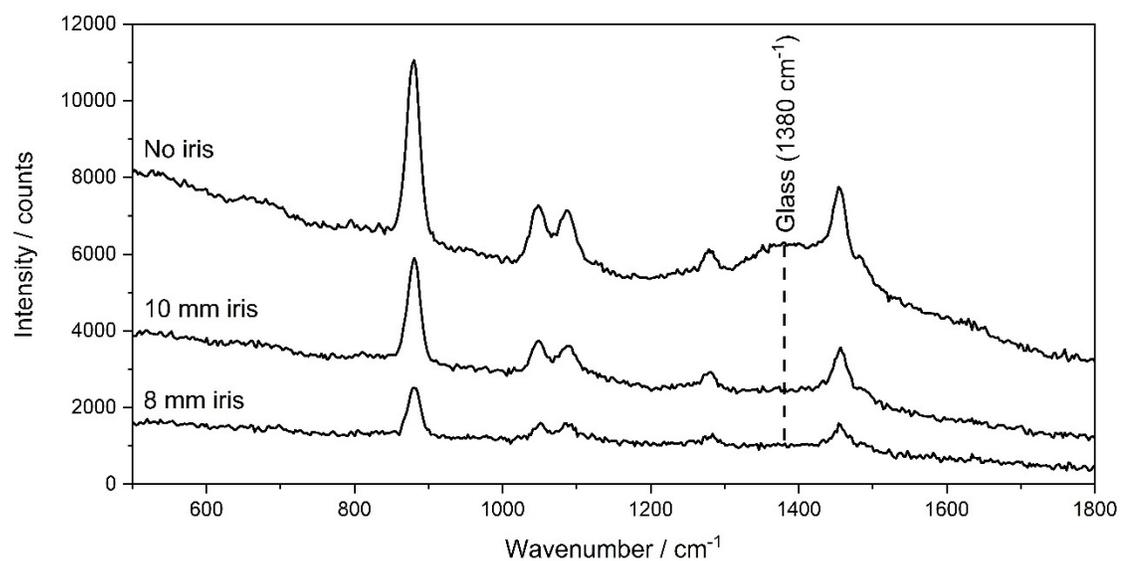

Fig. S9. Focus-matched ISORS spectra of a 70 mm diameter clear glass whisky bottle. Spectra are taken with an integration time of 5 s and 5 accumulations. Inclusion of an iris aperture of 10 mm successfully excludes the glass signal. Apertures of less than 10 mm also exclude the glass signal but the overall signal intensity is also reduced.